\documentclass{revtex4-2}
\usepackage{graphicx}
\usepackage{dcolumn}
\usepackage{amsmath}
\usepackage[latin1]{inputenc}
\usepackage{graphicx, psfrag}
\usepackage{amssymb}
\usepackage[colorlinks=true, citecolor=blue, urlcolor = blue, linkcolor= blue, bookmarks=true]{hyperref}
\usepackage{float}
\usepackage{amsmath}
\usepackage{amsfonts}
\usepackage{dcolumn}
\usepackage{hyperref}
\usepackage{subfigure}
\usepackage{pgfplots}
\usepackage{epstopdf}

\makeatletter
\def\btt#1{\texttt{\@backslashchar#1}}
\DeclareRobustCommand\bblash{\btt{\@backslashchar}} \makeatother

\makeatletter
\def\btt#1{\texttt{\@backslashchar#1}}
\DeclareRobustCommand\bblash{\btt{\@backslashchar}} \makeatother

\pgfplotsset{compat=1.18}
\begin{document}	
\title[]{Thermodynamics of Kerr-Sen-AdS black holes in the restricted phase space}

\author{Md Sabir Ali} 
\email{alimd.sabir3@gmail.com} 
\affiliation{Lanzhou Center for Theoretical Physics $\&$ Key Laboratory of Theoretical Physics of Gansu Province, Lanzhou University, Lanzhou, Gansu 730000, China\\
Key Laboratory of Quantum Theory and Applications of MoE, Lanzhou University, Lanzhou, Gansu 730000, China\\
Institute of Theoretical Physics $\&$ Research Center of Gravitation, Lanzhou University, Lanzhou 730000, China}
\author{Sushant G Ghosh}
\email{sghosh2@jmi.ac.in} 
\affiliation{Centre for Theoretical Physics, Jamia Millia
	Islamia, New Delhi 110025, India}
\affiliation{Astrophysics and Cosmology
	Research Unit, School of Mathematics, Statistics and Computer Science, University of
	KwaZulu-Natal, Private Bag 54001, Durban 4000, South Africa}
\author{Anzhong Wang}
\email{anzhong\_wang@baylor.edu} 
\affiliation{GCAP-CASPER, Physics Department, Baylor University, Waco, Texas 76798-7316, USA}
 \affiliation{Institute for Theoretical Physics and Cosmology, Zhejiang University of Technology, Hangzhou, 310032, China}
\begin{abstract}
\noindent	
We analyse the restricted phase space thermodynamics (RPST) of Kerr-Sen-AdS black holes with the central charge $C$ and its conjugate chemical potential $\mu$ but exclude the familiar $PdV$ term in the first law of black hole thermodynamics. That gives rise to a new perspective on the thermodynamics of black holes. Using the scaling properties, we investigate the first law and the corresponding Euler formula. Such formalism has its beauty, to say, for example, the mass is considered to be a homogeneous function of the extensive variables in the first order. In contrast, the intensive variables are of zeroth order. Because of the complicated expressions of the metric, we numerically calculate the critical values of the thermodynamic quantities. We find the phase transition behaviour of the free energy and other thermodynamic conjugate variables that appear in the first law. The RPST of the Kerr-Sen-AdS black holes is like that of the Reissner-Nordstr$\ddot{o}$m-AdS and the Kerr-AdS black holes. Such notions of the phase transition behaviour show that there should be some underlying universality in the RPST formalism.
\end{abstract}

\maketitle

\section{Introduction\label{Intro}}
Almost five decades after Hawking's proposal for the black hole as radiating black body \cite{Hawking}, the investigation of black hole thermodynamics is a significant area of research in modern theoretical physics. Since its inception by the pioneers Bardeen, Carter, and Hawking \cite{Bardeen, Hawking} of the laws of black hole mechanics, the thermodynamics of black holes have been an active and crucial subject area of investigation. However, Bekenstein's entropy-area relation \cite{Bekenstein:1973ur}, is a universal feature that is explored in many approaches and provides a well-posed interpretation of the counting of the microstate \cite{Jacobson:1993vj, Ashtekar:1997yu, Solodukhin:2011gn}. These interpretations on the microstate-counting give us the same entropy-area relation, $S=A/4$, in Einstein's general relativity. In black hole thermodynamics, the phase transition is a fascinating phenomenon that has received significant attention. It can be traced back to the pioneering research of phase transitions of asymptotically anti-de Sitter (AdS) black holes when Hawking and Page discovered a phase transition between the Schwarzschild AdS black hole and pure thermal AdS space \cite{Hawking2}. Thermodynamics black holes in AdS spacetime provide speculative ideas related to the ordinary thermal system and have a deep-rooted connection to the unresolved puzzles of quantum gravity. It also has a dual description of the boundary Conformal Field Theory (CFT) at finite temperature in AdS/CFT correspondence. The Hawking-Page phase transition relates the confinement/de-confinement phases in the double quark-gluon plasma \cite{Witten:1998qj}. For charged or rotating black holes, one likewise observes a small/large black hole first-order phase transition reminiscent of the liquid/gas transition of the van der Waals fluid. 

An exciting development is the identification of negative cosmological constant, $\Lambda (<0)$ as a positive thermodynamic pressure via $P=-\Lambda/8\pi G$, where $\Lambda=-{(d-1)(d-2)}/{2\ell^2}$, $d$ is the number of spacetime dimension and $\ell$ is the AdS curvature radius. It is traditionally called the extended phase space thermodynamics \cite{Kastor}, after introducing a new $(P, V)$ pair of state variables. After this development in the extended phase space, significant attention was devoted to Refs.~\cite{Dolan,Ali:2019myr, Ali:2019mkl, Dolan2,Dolan3,Kubiznak,Cai,Kubiznak2,Xu,Xu2,Zhm,Mir:2019ecg,Astefanesei:2019ehu, Tzikas:2018cvs,Kumar:2020cve} in the field. Such notion of the phase transitions behaviour led to the holographic deployment of black holes as systems dual to conformal field theories, quantum chromodynamics, and condensed matter physics. Another exciting feature that is included in connection to extended phase thermodynamics is the development of the correlation between the null geodesics and thermodynamic phase transitions.

Studying thermodynamics in the extended phase space has become one cornerstone of black hole mechanics and gave birth to conventionally called black hole chemistry \cite{Kubiznak2,Mir:2019ecg,Astefanesei:2019ehu, Tzikas:2018cvs,Kumar:2020cve}. Since its inception, a series of development has been made, and the most recent variant of the extended phase space thermodynamics is designed and investigated in Visser's formalism \cite{Visser}. Such investigations include considerations about the famous work of AdS/CFT correspondence by Maldacena \cite{Maldacena}. The main contribution of Visser's \cite{Visser} formalism includes the CFT central charge $C$, and its conjugate chemical potential $\mu$ in the first law of black hole thermodynamics. Here, we expect that the volume and the corresponding conjugate pressure
are transformed into the language of the AdS/CFT correspondence, i.e. $\mathcal{V}\sim L^{d-2}$ as a CFT volume in which $L$ is representing the familiar AdS radius, and the corresponding pressure term, $\mathcal{P}$, is
related through the CFT equation of states $E=(d-2)\mathcal{PV}$, where $d$ is the spacetime dimensions in the bulk. It is noticeable here that the very idea of inclusion of the chemical potential, $\mu$ and the corresponding central charge, $C$ as a pair of thermodynamic variables, have been considered in many of the previous works
\cite{Kastor2,Zhang,Karch,Maity,Wei}. Moreover, in the new formalism of Visser, the pairs $(P,V)$ variables are transformed into the $(\mathcal{P,V})$, where the AdS radius plays the fundamental role. From these considerations, together with Visser's formalism, we have for the charged rotating AdS black holes, the first law has the form
$d E = Td S -\mathcal{P} d\mathcal{V}+\tilde \Phi d\tilde Q+\Omega d J +\mu d C,$
which in turn gives rise to the Euler-like relation 
\begin{align}
E=TS +\tilde \Phi\tilde Q+\Omega J+\mu C,
\label{Elike}
\end{align}
where $\tilde Q$ is the rescaled electric charge and $\tilde \Phi$ is its conjugate potential. In Visser's framework, we have the full description of the thermodynamic
properties of the black hole in the bulk and the corresponding CFT description on the boundary. Hence there is a one-to-one correspondence in this grand formalism. The problem arrives somewhere else such that when one replaces the famous $(P, V)$ term with the CFT pairs $(\mathcal{P, V})$, interpreting the total mass as the black hole enthalpy is completely avoided. However, with the fixed central charge of $C$, Visser provides a thermodynamic description for the CFT which 
is holographically dual to the AdS black hole in the bulk \cite{CKM, Rafiee:2021hyj}. 

The lack of correct homogeneity relation of the internal energy and the intensive variables, extended phase space thermodynamics, and Visser's original formalism has the problem of an ``ensemble of theories". In the former case, we do not get the proper interpretation of the volume term in the first law or the Euler-like or Euler-Gibbs-Duhem relation. In contrast, for later case, there is a  missing $\mathcal{P}\mathcal{V}$ term in the Euler-like relation \eqref{Elike}. Such issues could be circumvented if one assumes a restricted version of Visser's formalism called the {\em restricted phase space thermodynamics} (RPST) for AdS black holes as proposed in \cite{Gao}. The RPST formalism has been studied for Reissner-N$\Ddot{o}$rdtsrom-AdS and Kerr-AdS black holes. Since the RPST formalism lacks the familiar ``PdV" term in the first law, it cannot give information about the black holes heat engine. Nevertheless, having been occupied by multiple degrees of freedom as still, it is the thermodynamic system, the thermodynamic behaviors in the RPST formalism provide us with other exciting phenomena. Following \cite{Gao:2021xtt, Gao}, in the present work, we study the RPST formalism for the case of Kerr-Sen-AdS black holes \cite{Wu:2020cgf}. It can be seen that,
despite the differences in their geometries, the thermodynamic properties of the Kerr-Sen-AdS black holes in RPST formalism mimic similar behaviors to that of the RN-AdS and Kerr-AdS. Such similarities suggest that there should be some universal features in the RPST formalism. Moreover, studying black hole thermodynamics 
in the RPST formalism may also be helpful in further understanding the  AdS/CFT correspondence.

The organization of the paper is as follows. In Sec.~\ref{kerr-sen-ads}, we briefly review the Kerr-Sen-AdS spacetime and derive the various thermodynamic quantities, the first law, and the corresponding Euler-like relation. In Sec.~\ref{EOS1}, we express the black hole mass in terms of extensive variables such as entropy $S$, the CFT charge $C$, the AdS radius $\ell$ the Dilaton-Axion charge $Q$, and the angular momentum $J$, and represent the intensive variables in terms of the equations of states. Such a process confirms the correct homogeneity behaviours of the extensive and intensive variables. Sec.~\ref{RPST1} is mainly devoted to investigate different thermodynamic processes. Finally, in Sec.~\ref{conclusions}, we summarize our results and conclude the paper.

\section{Kerr-Sen-AdS black hole and the restricted phase space thermodynamics}\label{kerr-sen-ads}
This section briefly reviews the Kerr--Sen black hole and its extension to the anti-de Sitter spaces. Sen \cite{Sen:1992ua} discovered a charged rotating black holes solution to the low-energy limit of the heterotic string theory, famously known as the Kerr--Sen black holes. The action is a modification to that of general relativity from that of the low-energy heterotic string theory, given by
\begin{equation}
\label{aksi}
S= \int d^4x \sqrt{-\tilde{g}} ~e^{-\Phi} \left[\mathcal{R} + (\nabla\Phi)^2 -\frac{1}{8} F^2-\frac{1}{12} H^2\right],
\end{equation}
where $ \tilde{g} $ is a determinant of metric tensor $ g_{\mu\nu} $, $ \mathcal{R} $ is a Ricci scalar, $F=F_{\mu\nu}F^{\mu\nu}$ with $F_{\mu\nu}$ being the $U(1)$ Maxwell field strength tensor, $ \Phi $ is a scalar dilaton field, and $H = H_{\mu \nu \rho} H^{\mu \nu \rho}$ is the field strength for the axion field
\begin{equation}\label{action2}
H_{\kappa\mu\nu} =\partial_{\kappa}B_{\mu\nu}+ \partial_{\nu}B_{\kappa\mu}+\partial_{\mu}B_{\nu\kappa}-\frac{1}{4}\left(A_{\kappa}F_{\mu\nu}+A_{\nu}F_{\kappa\mu}+A_{\mu}F_{\nu\kappa}\right).
\end{equation}
It is a 3-form tensor field which contains the antisymmetric 2-form tensor field where the last term in Eq.~(\ref{action2}) is the gauge Chern-Simons term. On using the conformal transformation 
\begin{equation}
\label{conformal}
ds_{E}^2 = e^{-\Phi} \tilde{ds}^2,
\end{equation}
one obtains the action in Einstein's frame
\begin{equation}
S= \int d^4 x \sqrt{-g} \left[R(g)-\frac{1}{2}(\nabla\Phi)^2 -\frac{e^{-\Phi}}{8}F^2-\frac{e^{-2\Phi}}{12}H^2\right],\label{EinAction}
\end{equation}
which encompasses the Einstein-Hilbert action as a particular case in the absence of dilaton, vector, and axion fields. \\

To obtain the black holes with a nonzero cosmological constant,  
one can rewrite the three-form field 
$H \equiv d\mathcal{B} -A\wedge\, F/4 = -e^{2\phi}\,{\star}d\chi$, where $\mathcal{B}$ is
an anti-symmetric two-form potential, and the $\star$ operator is the Hodge duality.
Then the Lagrangian can be rewritten in a different but completely equivalent form \cite{Wu:2020cgf}
\begin{eqnarray}
\hat{\mathcal{L}} &=& \sqrt{-g}\big[R -\frac{1}{2}(\partial\phi)^2 -\frac{1}{2}e^{2\phi}(\partial\chi)^2
 -e^{-\phi}F^2\big]  \nonumber \\
&&+\frac{\chi}{2}\epsilon^{\mu\nu\rho\lambda}F_{\mu\nu}F_{\rho\lambda} +  +\sqrt{-g}\big[4 +e^{-\phi} +e^{\phi}\big(1 +\chi^2\big)\big]/\ell^2 \, ,  
\end{eqnarray}
where $\epsilon^{\mu\nu\rho\lambda}$ is the four-dimensional Levi-Civita antisymmetric tensor
density and 
$\ell$ is the cosmological scale or the reciprocal of the gauge coupling constant. However, in the usual Boyer-Lindquist coordinate, the Kerr--Sen--AdS metric reads 
\begin{widetext}
\begin{align}
\mathrm {ds^{2}}=-\frac{\Delta_{r}}{\rho^{2}}\left(dt
-\frac{a \sin ^{2} \theta }{\Xi}d\phi\right)^{2}
+\frac{\rho^{2}}{\Delta_{r}} dr^{2}+\frac{\rho^{2}}{\Delta_{\theta}} d\theta^{2} 
+\frac{\sin ^{2} \theta \Delta_{\theta}}{\rho^{2}}\left(a dt
-\frac{r^{2}+2br+a^{2}}{\Xi} d\phi\right)^{2},
\end{align}
\end{widetext}

where
\begin{eqnarray*}
&& \rho^{2} =r^{2}+2 b r+a^{2} \cos ^{2} \theta,  \qquad \\
&& \Delta_{r} =\left(r^{2}+2br+a^{2}\right)\left(1+\frac{r^{2}+2br}{\ell^{2}}\right)-2 G m r, \\
&& \Xi =1-\frac{a^{2}}{\ell^{2}},  \qquad
\Delta_{\theta} =1-\frac{a^{2}}{\ell^{2}} \cos ^{2} \theta.
\end{eqnarray*}
The parameter $b$ is the diatonic charge of the black holes and is expressed as $b=q^2/(2m)$, where $q$ is the electric charge and $m$ is the mass of the black holes. In the limit of $\ell\to\infty$, the above metric reduces to the usual Kerr-Sen black holes. The non-rotating case ($a=0$), reduces to the Gibbons-Maeda-Garfinkle-Horowitz-Strominger (GMGHS) solution. Gibbons and Maeda \cite{Gibbons:1987ps} gave the black hole and black brane solutions to the Dilaton field, and Garfinkle-Horowitz-Strominger obtained its charged version \cite{Garfinkle:1990qj}. It is worthwhile to mention that the Kerr-Sen-(A)dS black holes in four dimensions have been studied from various perspectives, including the black hole silhouettes \cite{Zhang:2021wda} and the phase space thermodynamics in the extended phase space \cite{Sharif:2021yis}. The ADM mass $M$, the angular momentum $J$ and the charge $Q$ in AdS spacetimes is related as
\begin{eqnarray}
\label{Therm}
&&M = \frac{m}{\Xi^2} \, , \quad J = \frac{ma}{\Xi^2} \, , \quad
Q= \frac{q}{\Xi}
\end{eqnarray}
where the mass term $m$ can be derived from the condition $\Delta_r(r=r_+)=0$ with $r_+$ being the radius of the event horizon. Therefore, we can write the $M$ and $J$ in terms of $a,\;r_+,\;G$ as 
\begin{eqnarray}
\label{mJ}
&& M=\frac{1}{2\Xi^2 G r_+}\left(r_+^{2}+2br_++a^{2}\right)\left(1+\frac{r_+^{2}+2br_+}{\ell^{2}}\right),\; \\ && J=\frac{a\left(r_+^{2}+2br_++a^{2}\right)}{2\Xi^2 G r_+}\left(1+\frac{r_+^{2}+2br_+}{\ell^{2}}\right)
\end{eqnarray}
The corresponding expressions for Kerr-AdS black holes are obtained when $b=0$ \cite{Gao:2021xtt},
\begin{eqnarray}
\label{mK}
&& M=\frac{1}{2\Xi^2 G r_+}\left(r_+^2+a^2+\frac{r_+^{4}}{\ell^{2}}+\frac{a^2r_+^{2}}{\ell^{2}}\right),\; \\ && J=\frac{a}{2\Xi^2 G r_+}\left(r_+^2+a^2+\frac{r_+^{4}}{\ell^{2}}+\frac{a^2r_+^{2}}{\ell^{2}}\right)
\end{eqnarray}
The rest of the thermodynamic quantities can be written as
\begin{eqnarray}
\label{therm1}
&&T =\frac{a^2 \left(r_+^2-\ell^2\right)+r_+^2 \left(4 b^2+8 b r_++\ell^2+3 r_+^2\right)}{4 \pi  \ell^2 r_+ \left(a^2+2 b r_++r_+)\right)}\, , \\
&&S = \frac{\pi(r_+^2 +2br_+ +a^2)}{G\Xi} \, , \quad
\Omega = \frac{a\Xi}{r_+^2 +2br_+ +a^2}+\frac{a}{\ell^2}\, , \\
&&\Phi = \frac{qr_+}{r_+^2 +2br_+ +a^2} \, .
\end{eqnarray}
The above thermodynamic quantities reduce to the corresponding quantities of Kerr-AdS black holes, in the limit, when $b=0$  \cite{Gao:2021xtt}
\begin{eqnarray}
\label{therm1}
&&T =\frac{r_+}{4\pi(r_+^2+a^2)}\left(1+\frac{a^2}{\ell^2}+\frac{3r_+^2}{\ell^2}-\frac{a^2}{r_+^2}\right)\, , \\
&&S = \frac{\pi(r_+^2+a^2)}{G\Xi} \, , \quad
\Omega = \frac{a\Xi}{r_+^2+ +a^2}+\frac{a}{\ell^2},
\end{eqnarray}
The contribution of $ \Phi $ apparently is non-vanishing when $ b=0 $. However, if we carefully analyse the relation $b=q^2/(2m)$, the potential term is trivially vanishing.
The above expressions are a simple derivation of the usual thermodynamics of the black hole systems. This work aims to introduce the RPST formalism studied in the RN-AdS and Kerr-AdS black holes \cite{Gao, Gao:2021xtt}. Proceeding with the same line of thought, we introduce the pairs for which we account RPST formalism. We defined such variables as \cite{Visser, Gao, Gao:2021xtt}
\begin{align}
C &=\frac{\ell^2}{G},\quad
\mu= \frac{M-TS-\Omega J-\hat{\Phi}{\hat{Q}}}{C},
\label{Cmu}
\end{align}	
where $\hat{Q}$ is the rescaled electric charge and $\hat{\Phi}$ is the corresponding rescaled electric potential, which is expressed as
\begin{eqnarray}
\label{rescaled_phi}
\hat{Q}=\frac{Q\ell}{\sqrt{G}},\;\;\hat{\Phi}=\frac{\Phi\sqrt{G}}{\ell}
\end{eqnarray}
Noticeably, all other thermodynamic quantities must always be positive except for the chemical potential, $\mu$. 
One should remember that we can also identify the chemical potential as an independent
thermodynamic quantity using the language of AdS/CFT only, as
$Z_{\rm CFT}=Z_{\rm Gravity}$, and using the expression for
the Gibbs free energy $W=\mu C=-T\log Z_{\rm CFT}=-T\log Z_{\rm Gravity}$, with 
$Z_{\rm Gravity}=\exp(-\mathcal{A}_E/T)$, $\mathcal{A}_E$ representing the
Euclidean action evaluated at the black hole event horizon (see 
\cite{Gibbons,Chamblin,Gibbons2} and references therein for details).
The RPST formalism differs from Visser's AdS/CFT duality description. In the formal case, we treat the AdS radius $\ell$, as a constant, and therefore its variation will vanish. Keeping this in mind and using the thermodynamic quantities as derived in Eqs.~(\ref{mJ}) and (\ref{therm1}), we see that the first law of thermodynamics is easily written as
\begin{align}
d M=Td S+\Omega d J+\hat{\Phi}d\hat{Q}+\mu d C,
\label{1st}
\end{align}
which is obviously free from the familiar $\mathcal{P}d\mathcal{V}$ term
and, hence Eq.~(\ref{Cmu}) immediately follows the Euler-like relation as,
\begin{align}
M=TS+\Omega J+\hat{\Phi}\hat{Q}+\mu C
\label{euler}
\end{align}
We should emphasize that Eqs.~\eqref{euler} and \eqref{1st} are fundamental in the RPST's formalism. These two equations inherently determine the probable Hawking-Page-like page transitions, as will be clear from the subsequent discussions. \\

As mentioned in \cite{Gao, Gao:2021xtt}, one should remember that although the pair $(\mu,C)$ is borrowed from the CFT dictionary, they could be understood as the chemical potential corresponding to the effective $N_{\rm bulk}$ relating the microscopic degrees of freedom defined in bulk. We should realize this as $\mu_{\rm CFT}=\mu_{\rm bulk}$, and $C=N_{\rm bulk}$ in the language of AdS/CFT dictionary. Therefore, for simplicity of the notations, we replace the pair $(\mu_{\rm bulk},N_{\rm bulk})$ of the bulk to  $(\mu,C)$. One should keep track of the fact that we study the black holes thermodynamics only at the bulk. It is also emphasized here that the pair $(\mu, C)$ is followed by the rescalings: $\mu\to\lambda^{-1}\mu$ and $C\to\lambda C$, where $\lambda$ is an arbitrary and non-zero constant. We also remembered it with such scaling, the first law and subsequent Euler-like relation do not get harmed.\\

\section{Equations of states and homogeneity}
\label{EOS1}
In this section, we are determined to express the previously derived thermodynamic quantities, e.g., the mass $M$, as well as other thermodynamic variables $T,\Omega,\hat{\Phi}, \mu$ relating the $S,J,\hat{Q}, C$ as extensive variables. 
We can circumvent this in the subsequent steps as follows. Moreover, one has to rewrite $a,G$ in terms of 
$(J,M,C)$, 
\begin{align}
a &=\frac{J}{M},\quad 
G=\frac{\ell^2}{C}.
\label{aG}
\end{align}
As we know, we may also solve the horizon radius $r_+$, in terms of entropy $S$, mass $M$, the central charge $C$ and the rescaled charge $\hat{Q}$ as 
\begin{align}
r_+=\frac{1}{{2 M}}\left({\sqrt{\frac{-4 J^2 (\pi  C+S)+\pi  C Q^4+4 \ell^2 M^2 S}{\pi  C}}-Q^2}\right).
\label{r2}
\end{align}
The case of Kerr-AdS black holes is derived by simply putting $Q=0$,
\begin{align}
r_+=\frac{1}{{2 M}}\left({\sqrt{\frac{-4 J^2 (\pi  C+S)+4 \ell^2 M^2 S}{\pi  C}}}\right).
\label{r2}
\end{align}
The horizon radius for Gibbons-Maeda- Garfinkle-Horowitz-Strominger (GMGHS) black holes in AdS spacetimes is written as
\begin{align}
r_+=\frac{1}{{2 M}}\left({\sqrt{\frac{\pi  C Q^4+4 \ell^2 M^2 S}{\pi  C}}-Q^2}\right).
\label{r2}
\end{align}
Using Eqs.\,\eqref{aG} and \eqref{r2} into Eq.\eqref{Therm}, we can express the ADM mass of the Kerr-Sen-AdS black holes in terms of extensive variables as 
\begin{align}
M=\frac{\sqrt{\pi  C+S} \sqrt{\pi  C \left(4 \pi ^2 J^2+S^2\right)+2 \pi ^2 \hat{Q}^2 S+S^3}}{2 \ell\pi ^{3/2} \sqrt{C} \sqrt{S}}.
\label{MSJC}
\end{align}
All other thermodynamic variables are conventionally derived using the first law, Eq.~(\ref{1st}).  
Using Eq.\,\eqref{MSJC} and the first law, the followings are the equation of states 
\begin{align}
&T=\left(\frac{\partial M}{\partial S}\right)_{C,\hat{Q}, J}=\frac{C^2 \left(\pi ^2 S^2-4 \pi ^4 J^2\right)+4 \pi  C S^3+2 \pi ^2 \hat{Q}^2 S^2+3 S^4}{4 \pi ^{3/2} \ell\sqrt{C} S^{3/2} \sqrt{\pi  C+S} \sqrt{\pi  C \left(4 \pi ^2 J^2+S^2\right)+2 \pi ^2 \hat{Q}^2 S+S^3}}, 
\label{TSJC}\\
&\Omega=\left(\frac{\partial M}{\partial J}\right)_{C,S, \hat{Q}}
=\frac{2 \pi ^{3/2} \sqrt{C} J \sqrt{\pi  C+S}}{\ell \sqrt{S} \sqrt{\pi  C \left(4 \pi ^2 J^2+S^2\right)+2 \pi ^2 \hat{Q}^2 S+S^3}}, 
\label{OmSJC}\\
&\hat{\Phi}=\left(\frac{\partial M}{\partial \hat{Q}}\right)_{C,S,J}=\frac{\sqrt{\pi } \hat{Q} \sqrt{S} \sqrt{\pi  C+S}}{\ell\sqrt{C}  \sqrt{\pi  C \left(4 \pi ^2 J^2+S^2\right)+2 \pi ^2 \hat{Q}^2 S+S^3}},
\label{PhSJC}\\
&\mu=\left(\frac{\partial M}{\partial C}\right)_{S, \hat{Q},J}=\frac{\pi ^2 C^2 \left(4 \pi ^2 J^2+S^2\right)-2 \pi ^2 \hat{Q}^2 S^2-S^4}{4 \pi ^{3/2} \ell C^{3/2} \sqrt{S} \sqrt{\pi  C+S} \sqrt{\pi  C \left(4 \pi ^2 J^2+S^2\right)+2 \pi ^2 \hat{Q}^2 S+S^3}}.
\label{muSJC}
\end{align}
The above thermodynamic quantities reduce to that of the GMGHS-AdS black holes thermodynamic expressions such that 
\begin{align}
M=\frac{\sqrt{\pi  C+S} \sqrt{\pi CS+2 \pi ^2 \hat{Q}^2+S^2}}{2 \ell\pi ^{3/2} \sqrt{C}}.
\label{MSJC}
\end{align}
\begin{align}
&T=\left(\frac{\partial M}{\partial S}\right)_{C,\hat{Q}}=\frac{\pi^2C^2+4 \pi  C S+2 \pi ^2 \hat{Q}^2+3 S^2}{4 \pi ^{3/2} \ell\sqrt{C} S^{1/2} \sqrt{\pi C+S} \sqrt{\pi  C S+2 \pi ^2 \hat{Q}^2+S^2}}, 
\label{TSJC}\\
&\hat{\Phi}=\left(\frac{\partial M}{\partial \hat{Q}}\right)_{C,S}=\frac{\sqrt{\pi } \hat{Q} \sqrt{\pi  C+S}}{\ell\sqrt{C}  \sqrt{\pi  C S+2 \pi ^2 \hat{Q}^2+S^2}},
\label{PhSJC}\\
&\mu=\left(\frac{\partial M}{\partial C}\right)_{S, \hat{Q}}=\frac{\pi^2 C^2 S -2 \pi^2 \hat{Q}^2 S-S^3}{4 \pi^{3/2} \ell C^{3/2} \sqrt{S} \sqrt{\pi C+S} \sqrt{\pi  C S+2 \pi ^2 \hat{Q}^2+S^2}}.
\label{muSJC}
\end{align}

These equations show that all the intensive variables, such as the mass $M$, the angular velocity $\Omega$, the rescaled electric potential $\hat{\Phi}$, and the chemical potential $\mu$  as the functions of $S,J,C, \hat{Q}$. Obviously, the ADM mass $M$ is scaled as $M\to\lambda M$, 
while $T,\Omega, \hat{\Phi},\;\text{and}\;\mu$ are not, with the proviso that the extensive variables are
scaled as $S\to \lambda S, J\to \lambda J, C\to \lambda C$ and $\hat{Q}\to\lambda \hat{Q}$. 
This tells us that $M$ is homogeneous in first order while $T,\Omega,\hat{\Phi},\mu$ are zeroth order in their homogeneity. Keeping all these arguments in mind and also the first law and corresponding Euler-like relation, given in Eqs.~(\ref{1st}) and (\ref{euler}), we can express the Gibbs-Duhem relation for chemical potential as
\[
d\mu= - \mathcal{S}d T- \mathcal{J}d\Omega-\hat{\mathcal{Q}}d\hat{\Phi},
\]
where $\mathcal{S}=S/C, \mathcal{J}=J/C$, and $\hat{\mathcal{Q}}=\hat{Q}/C$ are, respectively, the entropy per unit $C$-charge, the angular momentum per unit $C$-charge, and the rescaled electric charge per unit $C$-charge. It is to be mentioned that $\mathcal{S}, \mathcal{J}$, and $\hat{\mathcal{Q}}$ are homogeneous functions in $S,J,\hat{Q}, C$ with order zero. Therefore, the thumb rule for any standard thermodynamics system is that they all should follow the first law of black hole mechanics and the corresponding Euler and Gibbs-Duhem relations. Besides, all the intensive variables, including the internal energy, must follow the various homogeneity properties. In this sense, the RPST formalism must be valid because it follows the first law of black hole thermodynamics and the essential homogeneous properties of all the intensive variables, including the mass.

One can see from Eqs.\,\eqref{TSJC}-\eqref{muSJC}, that the algebraic relations stand for the eight state variables, but not all of them should be of equal importance to define the macro state of the system. However, for the Kerr-Sen-AdS black hole, one can look for the macro states which can be determined by only four of the eight variables, namely, $(T,S), (\Omega, J), (\hat{\Phi}, \hat{Q}), (\mu, C)$. The mass function in Eq.~(\ref{MSJC}) is function of five variables, Viz., the charge $\hat{Q}$, the angular momentum $J$, the central charge $C$, the entropy $S$, and the AdS radius $\ell$. Therefore, one must continuously change any of the five variables to study the system's thermodynamic behaviour. 

Let us make helpful comments before we go into the main discussion of the thermodynamic process. For any non-zero and positive definite temperature, the value of the angular momentum $J$,
\begin{align*}
J\leq J_{\rm max}=\frac{S \sqrt{\pi ^2 C^2+4 \pi  C S+2 \pi ^2 \hat{Q}^2+3 S^2}}{2 \pi ^2 C},
\end{align*}
must not exceed this value. We obtain the similar bound on the angular momentum parameters for the Kerr black holes as \cite{Gao:2021xtt}
\begin{align*}
J^{\rm Kerr}\leq J_{\rm max}^{\rm Kerr}=\frac{S \sqrt{\pi ^2 C^2+4 \pi  C S+3 S^2}}{2 \pi ^2 C},
\end{align*}
in terms of the macro state variables.
\section{Thermodynamic processes}\label{RPST1}
This section discusses the various thermodynamic processes for the Kerr-Sen-AdS black holes in the recently proposed restricted phase space analysis. As mentioned earlier, the equation of states \eqref{TSJC}-\eqref{muSJC} imposes four different constraints over the eight variables. As mentioned in the previous section, if one chooses any of the four pairs to be varied continuously, the macroscopic description occurs in the system under study. In all the expressions in the equation of states  \eqref{TSJC}-\eqref{muSJC}, the charge parameter $\hat{Q}$ was square, except in the expressions for the electric potential, $\hat{\Phi}$. Therefore, the parameters $\hat{Q}$ and $\hat{\Phi}$ would bear identical signatures. It drives us to take the positive values of the charge parameter, i.e., $\hat{Q}\geq 0$. One more thing to be mentioned here, since any of the equation of states depends explicitly on four extensive parameters, the investigations of the permissible macroscopic processes can be tough to analyze. We fix any two variables to avoid such complicacy and consider the macroscopic processes along simple curves.

Therefore, we restrict ourselves to some specific processes, such as $T-S$, $\Omega-J$, $\hat{\Phi}-\hat{Q}$ and $\mu-C$ 
processes. These are the conventional processes as they only incorporate one pair of canonically conjugate intensive-extensive variables. Although they may sometimes lead to complicated results and may even end in analyses where numerical techniques are used. It is essential to mention that for the Reissner-N$\ddot{o}$rdstrom-AdS, we can get the exact analysis, but the Kerr-AdS case requires the slow rotation limits to have the full view of the thermodynamic process. In our case we cannot have the result in analytic forms, hence we adopt the numerical methods in determining the thermodynamic processes.

\subsection{The critical points}
The first-order phase transition of the black holes, either in the extended phase space thermodynamics or in the RPST formalism, requires the analysis of the $T-S$ curves at fixed $J$. Such a curve tells us about the first-order phase transition below the critical points, but in the crucial points, they become second-order. In the RPST formalism, the $T-S$ diagram at fixed rescaled electric charge $\hat{Q}$ for the Reissner-N$\ddot{o}$rdstrom-AdS, and also the $T-S$ diagram for the Kerr-AdS black holes at fixed $J$ has been tested in the usual four-dimensional spacetimes. Our work extends the existing results to examine the effect of the RPST formalism in the thermodynamic process.

To analyze the critical points on the $T-S$ curve at fixed $J$ and $\hat{Q}$, we need to solve for the inflexion point, which is given by the following relations,
\begin{align}
\left(\frac{\partial{T}}{\partial{S}}\right)_{J,\hat{Q}, C}=0, \quad \left(\frac{\partial^2{T}}{\partial{S}^2}\right)_{J,\hat{Q},C}=0.
\label{cri-eq}
\end{align}
We use the EOS \eqref{TSJC}, to determine the critical points not by analytic method but in the numerical way. Therefore we can evaluate the critical points by their numerical fits. Before do it, let us verify the existing results in the slow rotation limits of the mass term and all the other intensive variables. Therefore, by series expanding the mass function $M=M(S,J,C)$ and also the other intensive variables in powers of $J$ up to second orders and for $\hat{Q}=0$, we get the equations,
\begin{eqnarray}
M(S,J,\hat{Q},C)&=&\frac{\left(\pi  C \left(2 \pi ^2 J^2+S^2\right)+S^3\right)}{2 \pi ^{3/2} \sqrt{C} \ell S^{3/2}},\\
T(S,J,\hat{Q},C)&=&\frac{S (\pi  C+S)^2 \left(-6 \pi ^3 C J^2+\pi  C S^2+3 S^3\right)}{4 \pi ^{3/2} \ell\sqrt{C} S^{5/2}},\\
\Omega(S,J,\hat{Q},C)&=&\frac{2 \pi ^{3/2} \sqrt{C} J }{\ell S^{3/2}},\nonumber\\
\mu(S,J,\hat{Q},C)&=&\frac{\pi S\left(2\pi^2 J^2+S^2\right)-S^3}{4\pi^{3/2}\ell \left(C S\right)^{3/2}}
\end{eqnarray}
All the above expressions match the corresponding results of the Kerr-AdS black holes. In the limit when $J\to 0$, these results gives rise to GMGHS black holes in AdS spacetimes
\begin{eqnarray}
\label{GMGHS}
&&M=\frac{\sqrt{\pi  C+S} \sqrt{\pi  C S+2 \pi ^2 \hat{Q}^2+S^2}}{2 \pi ^{3/2} \sqrt{C} \ell}\nonumber\\
&&T=\frac{\pi ^2 C^2+4 \pi  C S+2 \pi ^2 \hat{Q}^2+3 S^2}{4 \pi ^{3/2} \sqrt{C} \ell \sqrt{\pi  C+S} \sqrt{\pi  C S+2 \pi ^2 \hat{Q}^2+S^2}}\nonumber\\
&&\Phi=\frac{\sqrt{\pi } \hat{Q} \sqrt{\pi  C+S}}{\sqrt{C} l \sqrt{\pi  C S+2 \pi ^2 \hat{Q}^2+S^2}}\nonumber\\
&&\mu=\frac{S \left(\pi ^2 \left(-C^2\right)+2 \pi ^2 \hat{Q}^2+S^2\right)}{4 \pi ^{3/2} C^{3/2} l \sqrt{\pi  C+S} \sqrt{\pi  C S+2 \pi ^2 \hat{Q}^2+S^2}}
\end{eqnarray}
The mass and other equations of states for the GMGHS-AdS black holes are obtained this way. However, as expected, such results at the second order of the angular momentum $J$ must hold the first law, the corresponding Euler relations, and the Gibbs-Duhem relations. \\  

Using eq.\,\eqref{TSJC}, we can numerically solve the approximate critical point equations \eqref{cri-eq}.  

We find the solution of the numerical critical-points for the Kerr-Sen AdS black hole by dimensional analysis:
\begin{eqnarray}
S_c=k_1(\epsilon)\cdot C,~~
\hat{Q}=k_2(\epsilon)\cdot C,~~
\end{eqnarray}
where, $\epsilon=J/\hat{Q}$ is the angular momentum to charge ratio as a dimension independent variable. Such scaling yields the critical parameters
\begin{eqnarray}
\label{scjc}
k_1(\epsilon)&=&\frac{-55.7964 \epsilon ^6+313.683 \epsilon ^5+376.682 \epsilon ^4+179.553 \epsilon ^3+41.4088 \epsilon ^2+3.59711 \epsilon +0.049029}{-1.94285 \epsilon ^5+11.2937 \epsilon ^4+10.9136 \epsilon ^3+4.43796 \epsilon ^2+0.784196 \epsilon +0.0375307}\nonumber\\
S_c &=&\frac{C}{k_2(\epsilon)}\frac{-55.7964 \epsilon ^6+313.683 \epsilon ^5+376.682 \epsilon ^4+179.553 \epsilon ^3+41.4088 \epsilon ^2+3.59711 \epsilon +0.049029}{-1.94285 \epsilon ^5+11.2937 \epsilon ^4+10.9136 \epsilon ^3+4.43796 \epsilon ^2+0.784196 \epsilon +0.0375307},\nonumber\\
k_2(\epsilon) &=&\frac{-1250.47 \epsilon ^6+3668.25 \epsilon ^5+21809.1 \epsilon ^4+21620. \epsilon ^3+7916.38 \epsilon ^2+1086.96 \epsilon +39.2484}{-29.9273 \epsilon ^5+99.2271 \epsilon ^4+483.749 \epsilon ^3+333.596 \epsilon ^2+66.5065 \epsilon +3.06935},\nonumber\\
J_c&=&\frac{C \left(-29.9273 \epsilon ^6+99.2271 \epsilon ^5+483.749 \epsilon ^4+333.596 \epsilon ^3+66.5065 \epsilon ^2+3.06935 \epsilon \right)}{-1250.47 \epsilon ^6+3668.25 \epsilon ^5+21809.1 \epsilon ^4+21620. \epsilon ^3+7916.38 \epsilon ^2+1086.96 \epsilon +39.2484}
\end{eqnarray}
\begin{figure}
    \centering
\includegraphics[width=.48\textwidth]{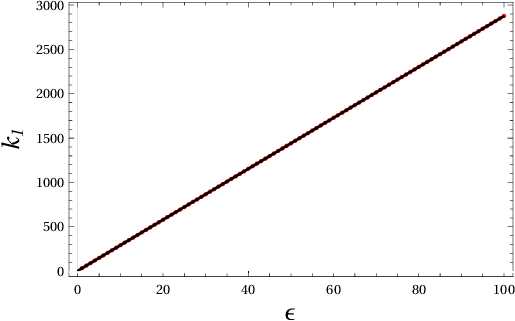}\hspace{2pt}
\includegraphics[width=.48\textwidth]{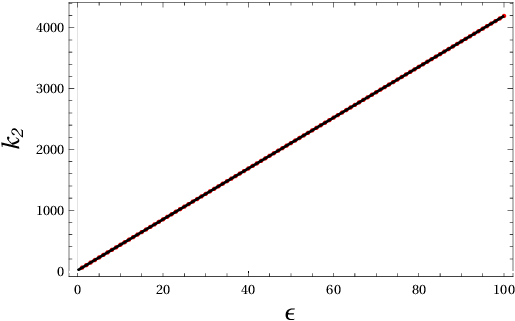}
    \caption{The plots for $k_1$ (left) and $k_2$ (right) vs the parameter $\epsilon$.}
    \label{fig:my_label}
\end{figure}
We choose to make the entropy, the temperature and the Helmholtz free energy dimensionless. The coefficients $k_1$ and $k_2$ as a function of $\epsilon$ are obtained by numerically solving the equations (36). They are also plotted as a function of $\epsilon$ in Fig.~(1). 

Inserting all the critical points in the above equation, the critical temperature reads
\begin{align}
T_c=\frac{3 k_1^4 \epsilon ^4-4 \pi ^4 k_2^2 \epsilon ^4+4 \pi  k_1^3 k_2 \epsilon ^4+\pi ^2 k_1^2 k_2 \epsilon ^3 \left(k_2 \epsilon +2\right)}{4 \pi ^{3/2} \ell \left(k_1 \epsilon \right){}^{3/2} \sqrt{k_2 \epsilon } \sqrt{k_1 \epsilon +\pi  k_2 \epsilon } \sqrt{k_1^3 \epsilon ^3+4 \pi ^3 k_2 \epsilon ^3+\pi  k_1^2 k_2 \epsilon ^3+2 \pi ^2 k_1 k_2 \epsilon ^2}}.
\label{tc}
\end{align}
In the same line, introduce the Helmholtz free energy as
\[
F(T,J,\hat{Q},C)=M(T,J,\hat{Q},C)-TS,
\]
which, in turn, on putting the exact expressions for the mass, and temperature, takes the value
\begin{align}
F(T,J,\hat{Q},C)=\frac{C \left(\pi ^2 C^2 \left(12 \pi ^2 J^2+S^2\right)+8 \pi ^3 C J^2 S-S^4\right)+2 \pi ^2 \hat{Q}^2 S (2 \pi  C+S)}{4 \pi ^{3/2} C l \sqrt{S} \sqrt{\pi  C+S} \sqrt{\pi  C^2 \left(4 \pi ^2 J^2+S^2\right)+2 \pi ^2 \hat{Q}^2 S+S^3}},
\label{free-e}
\end{align}
where the entropy $S$ is determined from Eq.\,\eqref{TSJC}. 
At the critical point, the free energy reads
\begin{eqnarray}
F_c &=&\Bigg[\frac{\sqrt{k_1} \epsilon  \big(\pi ^3 k_2 k_1^3 \left(k_2^2+6 k_2+10 \epsilon ^2\right)+2 \pi ^4 k_2^2 k_1^2 \left(3 k_2+10 \epsilon ^2+2\right)}{4 \pi ^{3/2} k_2^{3/2} \sqrt{k_1+\pi  k_2} \left(k_1^3+\pi  \left(k_1+2 \pi \right) k_2 k_1\right){}^{3/2}}\nonumber\\
&+&\frac{2 \pi ^5 k_2^2 k_1 \left(k_2 \left(5 \epsilon ^2+4\right)+6 \epsilon ^2\right)+16 \pi ^6 k_2^3 \epsilon ^2-k_1^6-\pi  k_2 k_1^5+\pi ^2 k_2^2 k_1^4\big)}{4 \pi ^{3/2} k_2^{3/2} \sqrt{k_1+\pi  k_2} \left(k_1^3+\pi  \left(k_1+2 \pi \right) k_2 k_1\right){}^{3/2}}\Bigg]\frac{C}{ \ell}.
\label{fc}
\end{eqnarray}
Eqs.\,\eqref{scjc}, \eqref{tc}, and \eqref{fc} contain complicated expressions for the critical points of various intensive and extensive variables. At a large parameter $\epsilon$ value, the critical points should match with the corresponding expressions for the Kerr-AdS black holes. Therefore, at large, e.g., $\epsilon\to\infty$, the critical points read
\begin{align*}
&S_c\approx 0.68250 \,C, \quad \,\, \,\,J_c \approx 0.02411 \,C,\\
&T_c\approx 0.26939 \,\ell^{-1},\quad
F_c \approx 0.10556 \,\ell^{-1} C.
\end{align*}
At this stage, we are able to express the equation of states \eqref{TSJC} and the free energy 
in terms of the reduced parameters
\begin{align*}
s=\frac{S}{S_c},\quad j=\frac{J}{J_c},\quad \tau=\frac{T}{T_c},\quad f=\frac{F}{F_c}. 
\end{align*}
Using the critical points of various extensive and intensive variables, we write the reduced temperature and the free energy in terms of reduced parameters as 
\begin{align}
\tau(s,j)&=\frac{C \epsilon ^2 \left(-4 \pi ^4 C^3 j^2 \epsilon ^2+\pi ^2 k_2^2 k_1^2 s^2 \left(C^3+2 \hat{Q}^2\right)+3 C^2 k_2^2 k_1^4 s^4 \epsilon ^2+4 \pi  C^3 k_2^2 k_1^3 s^3 \epsilon \right)}{4\ell \pi ^{3/2} k_2^2 \left(C k_1 s \epsilon \right){}^{3/2} \sqrt{C \left(k_1 s \epsilon +\pi \right)} \sqrt{4 \pi ^3 C^4 j^2 \epsilon ^4+\pi  C k_1 k_2^2 s \epsilon  \left(C^3 k_1 s \epsilon +2 \pi  \hat{Q}^2\right)+C^3 k_1^3 k_2^2 s^3 \epsilon ^3}}\\
f(t,j)&= \frac{C \epsilon  \left(4 \pi ^3 k_1 s \left(2 C^3 j^2 \epsilon ^4+k_2^2 \hat{Q}^2\right)+12 \pi ^4 C^3 j^2 \epsilon ^3+\pi ^2 k_2^2 k_1^2 s^2 \epsilon  \left(C^3+2 \hat{Q}^2\right)-C^3 k_2^2 k_1^4 s^4 \epsilon ^3\right)}{4\ell\pi ^{3/2} k_2 \sqrt{C k_1 s \epsilon } \sqrt{C \left(k_1 s \epsilon +\pi \right)} \sqrt{4 \pi ^3 C^4 j^2 \epsilon ^4+\pi  C k_1 s \epsilon  \left(C^3 k_1 s \epsilon +2 \pi  \hat{Q}^2\right)+C^3 k_1^3 s^3 \epsilon ^3}}
\end{align}
Let us emphasize here that, regarding the reduced parameters, the equation of states and the free energy is not absolutely independent of the central
charge $C$ and the rescaled electric charge $\hat{Q}$. But it does not reflect their effects in the temperature vs entropy behaviour plot or the free energy vs the temperature plot. Such plots mimic the same thermodynamic behaviour of the RN-AdS or Kerr-AdS black holes. We attribute such phenomena to any black hole spacetimes with any such charges. Similarly, we observe the same physical phenomena for any generic thermal systems or in the extended phase of black hole mechanics. Such events are characterized by the  {\em the law of corresponding states}. We attribute the role of the central charge to the number of particles when we study extended-phase space thermodynamics.
\begin{figure}[ht]
\begin{center}
\includegraphics[width=.48\textwidth]{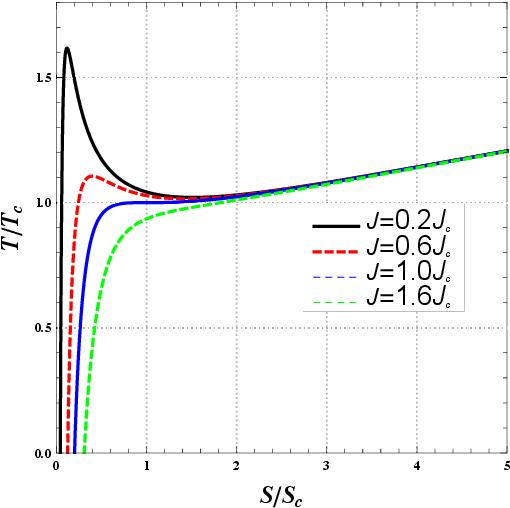}\hspace{2pt}
\includegraphics[width=.48\textwidth]{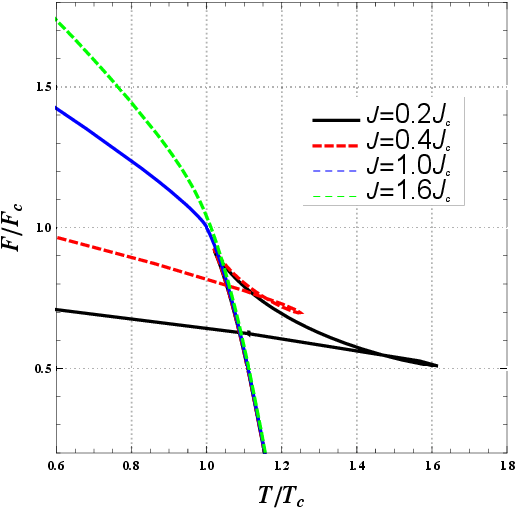}
\caption{$T-S$ and $F-T$ curves at fixed $J>0$}\label{fig1}
\end{center}
\end{figure}
We plot the $T/T_c-S/S_c$ and $F/F_c-T/T_c$ curves in the Fig.~\ref{fig1}at fixed $J$, where $J>0$. From the structure of the $T/T_c-S/S_c$ curves, there is clear an oscillatory phase below the critical points where $J<J_c$. Thus $T>T_c$ phase comprises three different black holes states, independent of angular momentum, the rescaled electric charge and the central charge at some particular $T=T_{0}$, where $T_0$ is the transition temperature. It corresponds to the first-order phase transition, where the small/large and metastable phases exist simultaneously. The small and large black hole phases are stable, while the metastable one is not. Therefore, the thermodynamic equilibrium exists between the small and large black hole phases. When $T_0=T_c$, such oscillatory behaviour and the metastable phase disappear and the corresponding phase becomes second order.\\
On a similar ground, if we look at the $F/F_c-T/T_c$ diagram, we see a swallowtail behaviour miming the first-order phase transition. When the temperature $T=T_0$, where the crossing of the swallow tail is reflected, is the first-order phase transition point. At the critical points, such behaviour ceases a point and represents the second-order phase transition point. Above this point, when $J>J_c$, the $F/F_c-T/T_c$ curve shows continuous behaviour. \\
Noticeably, we only deal with the cases where the angular momentum $J>0$. The cases of $J=0$ do not lead to the RN-AdS black holes phases for $T-S$ and $F-T$ curves for the Kerr-Sen AdS black holes. However, in this case, we have another interesting feature for the $T-S$ and $F-T$ diagrams. When $J=0$, we still have the stable small black hole phases, and the $T-S$ curve deals show similar behaviour as in the case of Kerr-Sen AdS black holes. Though the analytical evaluation of such minimum temperature for the Kerr-Sen-AdS black holes is impossible, luckily, we can get the minimum entropy for such a case. Therefore, using  Eq.\eqref{TSJC}, 
we get the minimum temperature is at the point when entropy has a value

\[
S_{\rm min}=\frac{1}{3} \pi \left(\sqrt{\frac{2 \left(C^3-9 \hat{Q}^2\right)}{\sqrt{\mathcal{A}}}+\mathcal{A}-6 \sqrt[3]{\frac{8 \hat{Q}^6}{C^3}-\hat{Q}^4}}+\sqrt{\mathcal{A}}-2 C\right), \mathcal{A}=3 \sqrt[3]{\frac{8 \hat{Q}^6}{C^3}-\hat{Q}^4}+C^2-\frac{6 \hat{Q}^2}{C}.
\]
Thus the value of $T_{\rm min}$ is obtained at $S=S_{\rm min}$, which is a very lengthy expression and is not of analytical interest. However, the rescaled equations of state in \eqref{TSJC} and the free energy expression \eqref{free-e} can be recast as
\begin{align}
\hat \tau = \tau({\hat{s},J,\hat{Q},C}),\quad
\hat f =f({\hat{s},J,\hat{Q},C}),
\label{tff}
\end{align}
where $\hat s= S/S_{\rm min},\, \hat \tau=T/T_{\rm min}$ and $\hat f = F/F_{\rm min}$. The plots for $T/T_{\rm min}-S/S_{\rm min}$ and $F/F_{\rm min}-T/T_{\rm min}$ are depicted in Fig.~\ref{fig1-1}. It is clear from the figure that for $T>T_{\rm min}$, there are two black hole phases at fixed values of $T, C,\;{\rm and}\; \hat{Q}$ but with different values of $S$. The smaller one corresponds to the unstable state, and the larger one corresponds to the stable state of the black holes. Apparently, there is no such metastable state for any $T>T_{\rm min}$ for such configurations.

\begin{figure}[ht]
\begin{center}
\includegraphics[width=.48\textwidth]{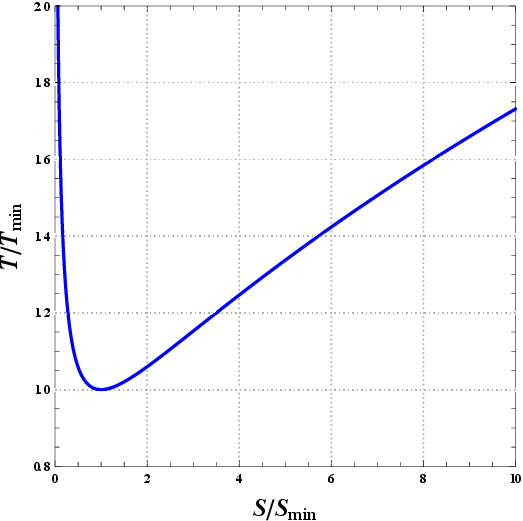}\hspace{2pt}
\includegraphics[width=.48\textwidth]{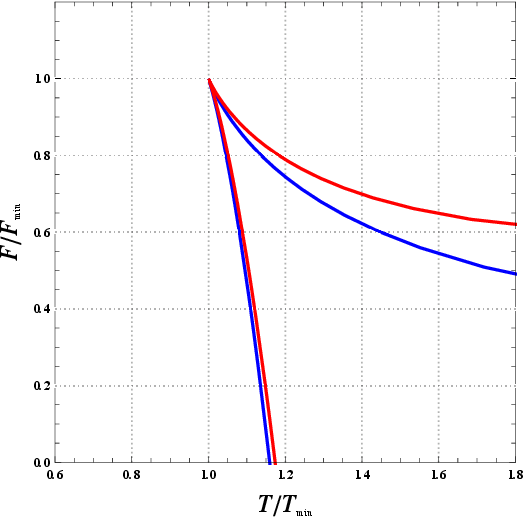}
\caption{$T-S$ and $F-T$ curves at $J=0$}\label{fig1-1}
\end{center}
\end{figure}
As a second example, we discuss the $T-S$ process at fixed values of $\Omega$. For that, we need to evaluate the functional dependence of $J$ in terms of $(\Omega)$ and putting them into \eqref{TSJC}, we find an expression  
$T=T(S,\Omega,\hat{Q},C)$. It also gives us a minimum for the temperature $T$, which is located at 
\begin{align*}
S_{\rm ex} =S_{\rm ex}(S,\Omega,\hat{Q},C),\quad
T_{\rm ex}=T_{\rm ex}(S,\Omega,\hat{Q},C).
\end{align*}
With the dimensionless variables 
$\tilde s=\frac{S}{S_{\rm ex}},\, \tilde \tau = \frac{T}{T_{\rm ex}}$, we can rewrite the equation of states as
the $T-S$ 
\begin{align}
\tilde \tau =\frac{\sqrt{\tilde{s} S_{\text{ex}}} \left(C \left(\tilde{s} S_{\text{ex}}+\pi  C\right) \left(\pi  C \tilde{s} S_{\text{ex}} \left(4-l^2 \Omega ^2\right)+3 \tilde{s}^2 S_{\text{ex}}^2+\pi ^2 C^2\right)+2 \pi ^2 \hat{Q}^2 \left(\tilde{s} S_{\text{ex}}+\pi  C \left(1-l^2 \Omega ^2\right)\right)\right)}{4\ell \pi ^{3/2} C \left(\tilde{s} S_{\text{ex}}+\pi  C\right) \sqrt{\tilde{s} S_{\text{ex}} \left(C \tilde{s} S_{\text{ex}} \left(\tilde{s} S_{\text{ex}}+\pi  C\right)+2 \pi ^2 \hat{Q}^2\right) \left(\tilde{s} S_{\text{ex}} \left(l^2 \Omega ^2+1\right)+\pi  C\right)}},
\label{tst}
\end{align}
It should be emphasized that as with $T/T_{\rm min}-S/S_{\rm min}$ given in \eqref{tff}, the behaviour of $T/T_{\rm ex}-S/S_{\rm ex}$ given in \eqref{tst}, are identically same.

For a clear view, we need to understand the 
$T-S$ processes at fixed $\Omega$. In addition, it is also important to see the effect of the $\mu-T$ relation $\mu=\mu(T,\Omega)$, where $\mu$ is served as the Gibbs free energy. Therefore, we replace $J$ with $\Omega$, in the $\mu-T$ relation. However, it is seen that $\mu$ is exactly peaked at $S=S_{\rm ex}$, where
\[
\mu_{\rm ex} =\mu_{\rm ex}(S, \Omega, \hat{Q}, C).
\]
Defining the dimensionless variable
\[
\tilde m =\frac{\mu}{\mu_{\rm ex}},
\]
the expected relation for $\mu-T$ can be written as
\begin{align}
\tilde m =\frac{\sqrt{\tilde{s} S_{\text{ex}}} \left(\tilde{s} S_{\text{ex}} \left(\tilde{s} S_{\text{ex}}+\pi  C\right) \left(\pi  C l^2 \Omega ^2 \tilde{s} S_{\text{ex}}-\tilde{s}^2 S_{\text{ex}}^2+\pi ^2 C^2\right)+2 \pi ^3 \hat{Q}^2 \left(\tilde{s} S_{\text{ex}} \left(l^2 \Omega ^2+1\right)+\pi  C\right)\right)}{4 \pi ^{3/2} C l \left(\tilde{s} S_{\text{ex}}+\pi  C\right) \sqrt{\tilde{s} S_{\text{ex}} \left(C \tilde{s} S_{\text{ex}} \left(\tilde{s} S_{\text{ex}}+\pi  C\right)+2 \pi ^2 \hat{Q}^2\right) \left(\tilde{s} S_{\text{ex}} \left(l^2 \Omega ^2+1\right)+\pi  C\right)}},
\label{mst}
\end{align}
where $\tilde s$ is determined from Eq.\,\eqref{tst} in terms of $\tilde \tau$, and the other variables.
It is worthy to mention that although Eqs.\,\eqref{tst} and \eqref{mst} are explicitly dependent on $\Omega$, this does not any significant effect on the $T/T_{\rm ex}$ vs $S/S_{rm ex}$ plots and also on the $\mu/\mu_{\rm ex}$ vs $T/T_{rm ex}$ plot.

\begin{figure}[ht]
\begin{center}
\includegraphics[width=.48\textwidth]{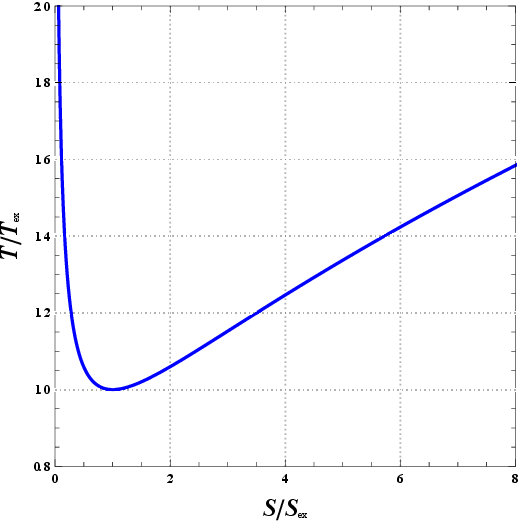}\hspace{2pt}
\includegraphics[width=.48\textwidth]{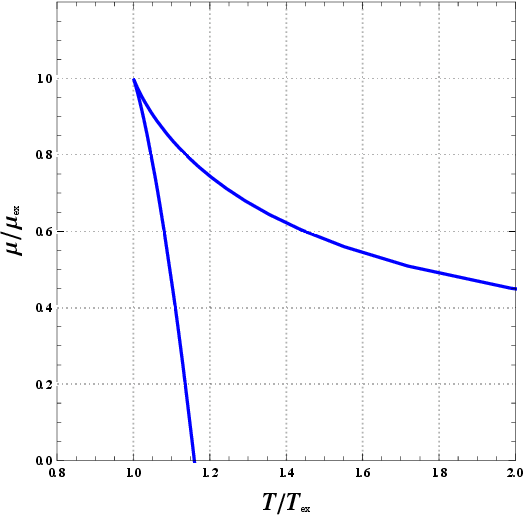}
\caption{$T-S$ and $\mu-T$ curves at fixed $\Omega$ }\label{fig1-2}
\end{center}
\end{figure}

Eqs.\,\eqref{tst} and \eqref{mst}, are depicted in Fig.~\ref{fig1-2}. 
The analytical expressions are not applicable even with a slow rotation limit, although we have shown it just for simplicity. We apply entirely numerical techniques to determine the analytical critical points. The corresponding numerical results do not change the notion of the plots. It seems to be the same as in the Kerr-AdS black holes case. 

\subsection{$\Omega-J$ process at fixed $S$}
Next, we emphasize on the $\Omega\;\text{vs}\; J$  processes at fixed values of $S$ and $\hat{Q}$. In the slow rotation limit, the expression for $\Omega$ is linearly dependant on $J$ at fixed values of $(S,\hat{Q},C)$ as seen from Eq.~({\color{blue}28}). Therefore, the corresponding curve 
is just a straight line in the slow rotation case for $\Omega-J$
curve (see Figure~{\textcolor{blue}{4}}). The graph for the 
the $\Omega-J$ processes is depicted in Figure~{\textcolor{blue}{4}}, where we take the relation  
\eqref{OmSJC}. Substituting $S = \mathcal{S} C$, $J = \mathcal{J} C$ and $\hat{Q}=\hat{\mathcal{Q}}/C$ into 
Eq.~\eqref{OmSJC}, we have 
\begin{align*}
\Omega=\frac{2 \pi ^{3/2} \mathcal{J} \sqrt{(\mathcal{S}+\pi )}}{\ell \sqrt{\mathcal{S}} \sqrt{\left(4 \pi ^3 \mathcal{J}^2+\mathcal{S}^2 (\mathcal{S}+\pi )+2 \pi ^2 \hat{\mathcal{Q}}^2 \mathcal{S}\right)}},
\end{align*}
where $\mathcal{S},\mathcal{J}$ both represent the intensive variables and is completely independent of the central charge $C$. We can see from the above expression that it is independent of the central charge $C$. In addition, we can determine the upper bound on the rescaled angular momentum
 $\mathcal{J}$, as 
\begin{align*}
\mathcal{J}_{\rm max}= 
\frac{\mathcal{S} \left(3 \mathcal{S}^2+4 \pi \mathcal{S}+\pi ^2+2 \pi ^2 \hat{\mathcal{Q}}^2\right)^{1/2}}
{2 \pi ^2}.
\end{align*}
The upper bound is determined to fix the endpoint of the value of the angular momentum in the $\Omega-J$ plot.

\begin{figure}[ht]
\begin{center}
\includegraphics[width=.4\textwidth]{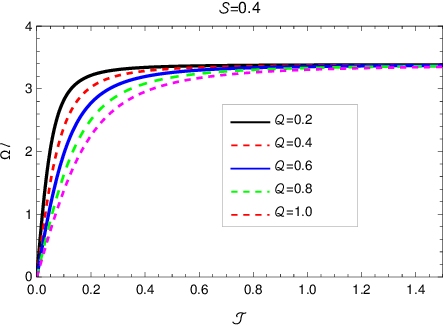}
\caption{$\Omega-\mathcal{J}$ curves at fixed $S$}\label{fig2}
\end{center}
\end{figure}

Figure \ref{fig2} depicts the behavior of the $\Omega-J$ curves at different values of $\mathcal{S}$ and $\hat{\mathcal{Q}}$. It is worth mentioning that the curves in the Figure.~\ref{fig2} are a monotonically varying functions. For smaller values of the angular momentum, the small segment of the $\Omega-J$ curve represents a straight line. It reaches a maximum and finally attains a fixed value at large values of the rescaled angular momentum. It is worth mentioning here that, unlike the Kerr-AdS case, the Kerr-Sen-AdS shows the maximum in the $\Omega-J$ plane. It signifies that a macroscopic phase transition behaviour of the small-large black hole phases should exist through this process.

\subsection{$\mu-C$ processes at fixed $(S,J)$}

We studied the $T-S$ and $\Omega-J$ processes in the previous subsection. In the present subsection, we see the behaviour of the $\mu-C$ processes. Such investigation is necessary because it might play an important role. For that purpose, we take the whole expression of the chemical potential in Eq.~\eqref{muSJC}. Remarkably, each $\mu-C$ curve at fixed $(S,\hat{Q},J)$, corresponds to a maximum. We obtain the maxima through the relation
\begin{align*}
C_{\rm max}=C_{\rm max}(S,\hat{Q},J),\quad
\mu_{\rm max}=\mu_{\rm max}(S,\hat{Q},J).
\end{align*}
for different values of the parameters $S$, $\hat{Q}$, and $J$. The value of the maximum chemical potential, $\mu_{\rm max}$ is always positive for any independent values of the parameters $S,\hat{Q},J$. We define the rescaled dimensionless quantities such that
\[
c=\frac{C}{C_{\rm max}},\quad m=\frac{\mu}{\mu_{\rm max}},
\]
therefore, the expression for the chemical potential is written as
\begin{align}
m=\frac{\pi ^2 c^2 C_{\max }^2 \left(4 \pi ^2 J^2+S^2\right)+2 \pi ^3 c \hat{Q}^2 S C_{\max }-S^4}{4 \pi ^{3/2} l \sqrt{S} \left(c C_{\max }\right){}^{3/2} \sqrt{\pi  c C_{\max }+S} \sqrt{\pi  c C_{\max } \left(4 \pi ^2 J^2+S^2\right)+2 \pi ^2 \hat{Q}^2 S+S^3}}.	
\label{mc}
\end{align}
in order to have a clear view of the graphs in the $\mu-C$ plots. It is important to mention that the expression for the chemical potential gives us a full understanding of the scaling properties of the  $\mu-C$ processes at any fixed values of $S$ and $J$. Besides that, it is important to mention that it also reflected a similar behaviour as with RN-AdS and Kerr-AdS black holes. This indicates that some universality lies behind the $\mu-C$ process for many black holes in AdS spacetimes. 
\begin{figure}[ht]
\begin{center}
\includegraphics[width=.4\textwidth]{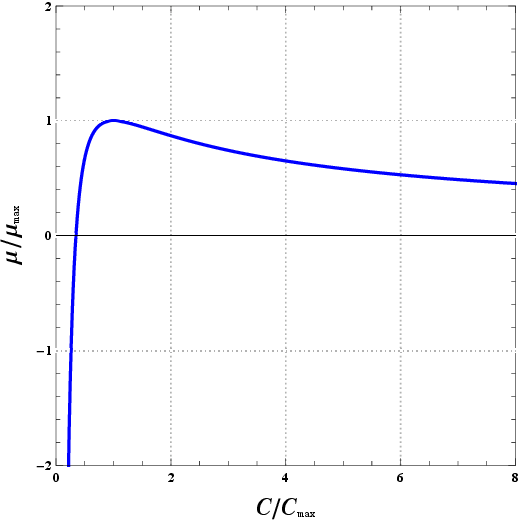}
\caption{$\mu-C$ curve at fixed $(S,J)$}
\label{fig3}
\end{center}
\end{figure}
The $\mu-C$ curve for the Kerr-Sen-AdS black holes is depicted in 
Fig.~\ref{fig3}. At some particular numerical value of the central charge, the chemical potential becomes zero, indicating a Hawking-Page-like phase transition.

\section{Conclusions}\label{conclusions}
The thermodynamics of the black hole in AdS spacetimes is of utmost importance because of its application to the AdS/CFT correspondence and related scenarios. This article presents thermodynamics from the viewpoint of central charge and its conjugate potential as a pair of new thermodynamic variables. In such a case,  we varied the gravitational constant to include it in the first law and the Euler or Euler-Gibbs-Duhem relation. The AdS radius is kept constant, unlike the case of extended phase space thermodynamics, where it is taken as the thermodynamic variable proportional to the pressure. 
\\
This paper discusses various thermodynamic processes of the AdS spacetimes in a fixed AdS radius while the gravitational constant is changed. Such a study is critical not only because it has the similar phase structure of the black holes as with varying cosmological constants but also because it has the dual description of the CFT, thereby showing some essential features of the AdS/CFT correspondence. In the grand canonical ensemble, the AdS/CFT correspondence relates the given macro state for the black holes in AdS spacetime to that of the dual CFT. 

We proceed by analysing the critical numerical values of the thermodynamic quantities. Such critical points lead us to find the exact nature of the temperature and the free energy of the AdS black holes for different values of the central charge $C$, the angular momentum $J$, and the charge parameter $\hat{Q}$. There is a certain minimum in the temperature vs the entropy profile showing a super-critical temperature where the free energy shows erratic behaviour. 

\section*{Acknowledgement}
 This work was supported by the National Natural Science Foundation of China
(Grants No. 11875151 and No. 12247101), the 111 Project under (Grant No. B20063) and
Lanzhou City's scientific research funding subsidy to Lanzhou University. S.G.G. would like to  thank SERB-DST for project No. CRG/2021/005771.


\begin{thebibliography}{99}

\bibitem{Hawking} S. W. Hawking, ``Particle Creation by Black Holes,'' 43 (1975) 199 [Erratum ibid. \textbf{46} (1976) 206]

\bibitem{Bardeen} J. M. Bardeen, B. Carter, and S. W. Hawking, ``The Four laws of black hole mechanics,''
Commun. Math. Phys. \textbf{31} (1973) 161

\bibitem{Bekenstein:1973ur}
J.~D.~Bekenstein,
``Black holes and entropy,''
Phys. Rev. D \textbf{7}, 2333 (1973)

\bibitem{Jacobson:1993vj}
T.~Jacobson, G.~Kang and R.~C.~Myers,
``On black hole entropy,''
Phys. Rev. D \textbf{49}, 6587-6598 (1994)

\bibitem{Ashtekar:1997yu}
A.~Ashtekar, J.~Baez, A.~Corichi and K.~Krasnov,
``Quantum geometry and black hole entropy,''
Phys. Rev. Lett. \textbf{80}, 904-907 (1998) 

\bibitem{Solodukhin:2011gn}
S.~N.~Solodukhin,
``Entanglement entropy of black holes,''
Living Rev. Rel. \textbf{14}, 8 (2011) 

\bibitem{Hawking2} S. W. Hawking, D.N. Page, ``Thermodynamics of black holes in 
anti-de Sitter space,'' Commun. Math. Phys. \textbf{87} (1983) 577

\bibitem{Witten:1998qj}
E.~Witten,
``Anti-de Sitter space and holography,''
 Adv. Theor. Math. Phys. \textbf{2}, (1998) 253.

\bibitem{Kastor} D. Kastor, S. Ray, J. Traschen, ``Enthalpy and the mechanics of AdS 
black holes,'' Class. Quant. Grav. \textbf{26}, 195011 (2009)

\bibitem{Dolan} B. P. Dolan, ``The cosmological constant and the black hole 
equation of state,'' Class. Quant. Grav. \textbf{28}, 125020 (2011)

\bibitem{Ali:2019myr}
M.~S.~Ali and S.~G.~Ghosh,
``Thermodynamics of rotating Bardeen black holes: Phase transitions and thermodynamics volume,'' Phys. Rev. D \textbf{99}, 024015 (2019)

\bibitem{Ali:2019mkl}
M.~S.~Ali, ``Ehrenfest scheme for $P$-$V$ criticality of the $d$-dimensional-AdS black holes surrounded by perfect fluid in Rastall theory,''
[arXiv:1901.04318 [gr-qc]]

\bibitem{Dolan2} B. P. Dolan, ``Pressure and volume in the first law of black
hole thermodynamics,'' Class. Quant. Grav. \textbf{28}, 235017 (2011)

\bibitem{Dolan3} Dolan, ``Compressibility of rotating black holes,'' Phys. Rev. D. \textbf{84} 127503 (2011)

\bibitem{Kubiznak} D. Kubiznak, R.B. Mann, ``P-V criticality of charged AdS black holes,''
JHEP 1207, \textbf{033} (2012)

\bibitem{Cai} R.~G. Cai, L.~M.~Cao, L.~Li, and R.~Q. Yang, ``P-V criticality in 
the extended phase space of Gauss-Bonnet black holes in AdS space,'' JHEP (2013) 005

\bibitem{Kubiznak2} D. Kubiznak, R. B. Mann, M. Teo, 
``Black hole chemistry: thermodynamics with Lambda,'' Class. Quantum Grav. \text{34} 063001 (2017)

\bibitem{Mir:2019ecg}
M.~Mir, R.~A.~Hennigar, J.~Ahmed and R.~B.~Mann,
``Black hole chemistry and holography in generalized quasi-topological gravity,''
JHEP \textbf{08}, 068 (2019)

\bibitem{Astefanesei:2019ehu}
D.~Astefanesei, R.~B.~Mann and R.~Rojas,``Hairy Black Hole Chemistry,''
JHEP \textbf{11}, 043 (2019)


\bibitem{Tzikas:2018cvs}
A.~G.~Tzikas,``Bardeen black hole chemistry,''
Phys. Lett. B \textbf{788}, 219-224 (2019). 

\bibitem{Kumar:2020cve}
A.~Kumar, S.~G.~Ghosh and S.~D.~Maharaj,``Nonsingular black hole chemistry,''
Phys. Dark Univ. \textbf{30}, 100634 (2020)

\bibitem{Xu} W. Xu, H. Xu, and L. Zhao, ``Gauss-bonnet coupling constant as a 
free thermodynamical variable and the associated criticality,'' Eur. Phys. J. C  \textbf{74} (2014) 2970

\bibitem{Xu2} W. Xu, L. Zhao, ``Critical phenomena of static charged AdS 
black holes in conformal gravity,''Phys. Lett. B \textbf{736} (2014) 214-220


\bibitem{Zhm}
M.~Zhang, D.~C.~Zou and R.~H.~Yue,
``Reentrant phase transitions and triple points of topological AdS black holes in Born-Infeld-massive gravity,''
Adv. High Energy Phys. \textbf{2017}, 3819246 (2017)

\bibitem{Hegde:2020xlv}
K.~Hegde, A.~Naveena Kumara, C.~L.~A.~Rizwan, A.~K.~M. and M.~S.~Ali,
``Thermodynamics, Phase Transition and Joule Thomson Expansion of novel 4-D Gauss Bonnet AdS Black Hole,''
[arXiv:2003.08778 [gr-qc]]

\bibitem{Hegde:2020yrd}
K.~Hegde, A.~Naveena Kumara, C.~L.~A.~Rizwan, M.~S.~Ali and K.~M.~Ajith,
``Null geodesics and thermodynamic phase transition of four-dimensional Gauss\textendash{}Bonnet AdS black hole,''
Annals Phys. \textbf{429}, 168461 (2021)

\bibitem{Hegde:2021hjq}
K.~Hegde, A.~Naveena Kumara, C.~L.~A.~Rizwan, M.~S.~Ali and K.~M.~Ajith,
``Thermodynamics, photon sphere and thermodynamic geometry of Eloy Ayon Beato -- Garcia Spacetime,''
[arXiv:2104.08091 [gr-qc]]

\bibitem{NaveenaKumara:2019nnt}
A.~Naveena Kumara, C.~L.~Ahmed Rizwan, S.~Punacha, K.~M.~Ajith and M.~S.~Ali,
``Photon orbits and thermodynamic phase transition of regular AdS black holes,''
Phys. Rev. D \textbf{102}, 084059 (2020)

\bibitem{Visser} M. R. Visser, ``Holographic thermodynamics requires a chemical 
potential for color,'' [\eprint{2101.04145}].

\bibitem{Maldacena} M. Maldacena, ``The large $N$ limit of superconformal field 
theories and supergravity,'' Adv. Theor. Math. Phys. \textbf{2} (1998) 231-252

\bibitem{Kastor2}
D.~Kastor, S.~Ray and J.~Traschen,
``Chemical Potential in the First Law for Holographic Entanglement Entropy,''
JHEP \textbf{11}, 120 (2014)

\bibitem{Zhang} 
J.~L.~Zhang, R.~G.~Cai and H.~Yu,
``Phase transition and thermodynamical geometry of Reissner-Nordstr\"om-AdS black holes in extended phase space,''
Phys. Rev. D \textbf{91}, 044028 (2015)

\bibitem{Karch} 
\bibitem{Karch:2015rpa}
A.~Karch and B.~Robinson,
``Holographic Black Hole Chemistry,''
JHEP \textbf{12}, 073 (2015)

\bibitem{Maity} 
R.~Maity, P.~Roy and T.~Sarkar,
``Black Hole Phase Transitions and the Chemical Potential,''
Phys. Lett. B \textbf{765}, 386-394 (2017)

\bibitem{Wei} 
S.~W.~Wei, B.~Liang and Y.~X.~Liu,
``Critical phenomena and chemical potential of a charged AdS black hole,''
Phys. Rev. D \textbf{96}, 124018 (2017)

\bibitem{CKM} 
W.~Cong, D.~Kubiznak and R.~B.~Mann,
``Thermodynamics of AdS Black Holes: Critical Behavior of the Central Charge,''
Phys. Rev. Lett. \textbf{127}, 091301 (2021)

\bibitem{Rafiee:2021hyj}
M.~Rafiee, S.~A.~H.~Mansoori, S.~W.~Wei and R.~B.~Mann,
``Universal criticality of thermodynamic geometry for boundary conformal field theories in gauge/gravity duality,''
Phys. Rev. D \textbf{105}, 024058 (2022)

\bibitem{Sen:1992ua}
A.~Sen,``Rotating charged black hole solution in heterotic string theory,''
Phys. Rev. Lett. \textbf{69}, 1006 (1992)


\bibitem{Gao} 
G.~Zeyuan and L.~Zhao,
``Restricted phase space thermodynamics for AdS black holes via holography,''
Class. Quant. Grav. \textbf{39}, 075019 (2022)

\bibitem{Gao:2021xtt}
Z.~Gao, X.~Kong and L.~Zhao,
``Thermodynamics of Kerr-AdS black holes in the restricted phase space,''
[\eprint{2112.08672}].

\bibitem{Wu:2020cgf}
D.~Wu, P.~Wu, H.~Yu and S.~Q.~Wu,
``Are ultraspinning Kerr-Sen- AdS$_4$ black holes always superentropic?,'' Phys. Rev. D \textbf{102}, 044007 (2020)

\bibitem{Gibbons:1987ps}
G.~W.~Gibbons and K.~i.~Maeda,
``Black Holes and Membranes in Higher Dimensional Theories with Dilaton Fields,''
Nucl. Phys. B \textbf{298}, 741 (1988)

\bibitem{Garfinkle:1990qj}
D.~Garfinkle, G.~T.~Horowitz and A.~Strominger,
``Charged black holes in string theory,''
Phys. Rev. D \textbf{43}, 3140 (1991)

\bibitem{Zhang:2021wda}
M.~Zhang and J.~Jiang,
``Strong cosmic censorship in near-extremal Kerr-Sen-de Sitter spacetime,''
 Eur. Phys. J. C \textbf{81}, 967 (2021)

\bibitem{Sharif:2021yis}
M.~Sharif and Q.~Ama-Tul-Mughani,
``$P-V$ criticality and phase transition of the Kerr-Sen-AdS Black Hole,'' Eur. Phys. J. Plus \textbf{136}, 284 (2021)

\bibitem{Wei2} 
S.~W.~Wei, P.~Cheng and Y.~X.~Liu,
``Analytical and exact critical phenomena of $d$-dimensional singly spinning Kerr-AdS black holes,''
Phys. Rev. D \textbf{93}, 084015 (2016)

\bibitem{Gibbons} 
G.~W.~Gibbons and S.~W.~Hawking,
``Action Integrals and Partition Functions in Quantum Gravity,''
Phys. Rev. D \textbf{15}, 2752-2756 (1977)

\bibitem{Chamblin} 
A.~Chamblin, R.~Emparan, C.~V.~Johnson and R.~C.~Myers,
``Charged AdS black holes and catastrophic holography,''
Phys. Rev. D \textbf{60}, 064018 (1999)

\bibitem{Gibbons2}
G.~W.~Gibbons, M.~J.~Perry and C.~N.~Pope,
``The First law of thermodynamics for Kerr-anti-de Sitter black holes,''
Class. Quant. Grav. \textbf{22}, 1503-1526 (2005)

\end{thebibliography}
\end{document}